\documentclass[preprint,showpacs,onecolumn]{revtex4}
\usepackage{graphicx}
\usepackage{dcolumn}
\usepackage{bm}
\usepackage{amssymb}
\usepackage{amsmath}
\usepackage{epsfig}
\usepackage{amssymb}
\usepackage{amsmath}
\usepackage{float}
\usepackage{amsfonts}
\usepackage{bbm}

\begin{document}

\title{Cross Kerr Effect Induced by Coupled Josephson Qubits in Circuit
Quantum Electrodynamics}
\author{Yong Hu}
\email{yhu3@ustc.edu.cn}
\affiliation{School of Physics, Huazhong University of Science and Technology, Wuhan,
Hubei 430074, China}
\author{Guo-Qin Ge}
\email{gqge@mail.hust.edu.cn}
\affiliation{School of Physics, Huazhong University of Science and Technology, Wuhan,
Hubei 430074, China}
\author{Shi Chen}
\affiliation{Department of Electronic Science and Technology, Huazhong University of
Science and Technology, Wuhan, Hubei 430074, China}
\author{Xiao-Fei Yang}
\affiliation{Department of Electronic Science and Technology, Huazhong University of
Science and Technology, Wuhan, Hubei 430074, China}
\author{You-Ling Chen}
\affiliation{Department of Physics, Peking University, Beijing 100871, China}

\begin{abstract}
We propose a scheme for implementing cross Kerr nonlinearity between two
superconducting transmission line resonators (TLR) via their interaction
with a coupler which is constructed by two superconducting charge qubits
connected to each other via a superconducting quantum interference device.
When suitably driven, the coupler can induce very strong cross phase
modulation (XPM) between the two TLRs due to its N-type level structure and
the consequent electromagnetically induced transparency in its lowest
states. The flexibility of our design can lead to various inter-TLR coupling
configurations. The obtained cross Kerr coefficient is large enough to allow
many important quantum operations in which only few photons are involved. We
further show that this scheme is very robust against the fluctuations in
solid state circuits. Our numerical calculations imply that the absorption
and dispersion resulted from the decoherence of the coupler are very small
compared with the strength of the proposed XPM.
\end{abstract}

\maketitle

\section{Introduction\label{Sec intro}}

The circuit quantum electrodynamics (QED) \cite{YalePRA,YaleNature} which
employs the superconducting coplanar transmission line resonator (TLR) to
substitute the standing-wave optical cavity and superconducting qubits \cite%
{SchonReview,DevoretReview,ClarkeReview} to replace the atoms is an on-chip
realization of cavity QED \cite{HarocheReview}. Compared with conventional
optical implementations, this solid-state architecture offers unprecedented
tunability and scalability which are leading to flexible quantum optics in
electronic circuits. Since the strong coupling between the TLRs and the
superconducting qubits with vacuum Rabi frequency up to three orders larger
than the qubit decoherence and cavity decay has already been achieved \cite%
{FrunzioIEEE,Transmon}, many important quantum information processes (QIP),
including coupling qubits using the TLR as a data bus \cite%
{QBusNIST,QBusYale} and preparing the TLR Fock states \cite%
{SinglePhoton,MultiPhotonNature,MultiPhotonPRL} have been demonstrated.

Stimulated by the advances on the single TLR level, recently the ideas of
photon manipulations between TLRs have been developed \cite%
{HuTLR,WilhelmTLR,SimmondsTLR}. The motivation is to facilitate the future
realization of scalable quantum computation. These schemes mainly consider a
model of two TLRs coupled to an assistant tunable coupler. One can tune on
and off the individual energy transfer between the TLRs and the coupler
through frequency selection and thus manipulate the two mode photon states.
A very latest experiment have realized the NOON state preparation in two
TLRs connected to an entanglement generator \cite{MartinisTLR}.

In optical systems, besides the linear tight-binding photon transfer, the
nonlinear Kerr interaction between cavity modes has also been studied
extensively. The cross Kerr nonlinearity, or the so-called cross phase
modulation (XPM), has found wide application in QIPs including the
construction of nontrivial quantum gates \cite{KimbleGate,RebicGate,XiaoGate}%
, the preparation of entangled photon states \cite{KerrPhoton,Xiaophoton},
and quantum non-demolition measurement \cite{Bell}. Enhancement of
dissipation-free photon-photon interactions at the few-photon level is a
fundamental challenge in quantum optics. Since photons can hardly interact
with each other, the XPM is often obtained by coupling two photon modes to
an atomic nonlinear media. To minimize the dispersion and absorption,
schemes of exploiting destructive quantum coherence in N type atoms have
been proposed \cite{NScheme} and realized \cite{NExp,NExTaiwan,NExShanxi}.
The reported experiments are performed in the semiclassical region due to
the very small interaction strength between the laser fields and $^{87}$Rb
atoms, which is often on the same level of the atomic decoherence rates. In
order to obtain significant phase shift,the probe and control pulses contain
large numbers of photons. The weak coupling places a major hindrance to the
further application of the Kerr effect on the single photon level.

In this paper, we aim at the realization of strong cross Kerr coupling
between two TLR modes. This work is inspired by the recent self phase
modulation scheme in circuit QED system \cite{RebicSPM}. We design a
superconducting circuit which exhibits complete analogy to the N level XPM
schemes in atomic systems. The proposed four level artificial molecule is
constructed by two Josephson charge qubits coupled by a superconducting
quantum interference device (SQUID). When capacitively coupled to the two
TLRs, the molecule induces cross Kerr interaction between microwave photons
in the two TLRs. The strong TLR-molecule coupling can boost the XPM strength
up to several $\mathrm{MHz}$. Moreover, the dispersion and absorption
resulted from the decoherence of the coupler are estimated to be negligible
compared with the obtained XPM strength. Our system is flexible enough to
allow various inter-TLR coupling configurations. Since long lifetimes for
both the TLRs and the molecule have already been achieved, many QIPs between
the TLRs in which there are only few photons involved can be performed with
very high fidelities.

The paper is organized in the following manner. We first briefly describe
the general N-type scheme of XPM in Sec. \ref{Sec theory} and then study its
realization in circuit QED system in Sec. \ref{Sec Natom}. In Sec. \ref{Sec
deco}, the influence of the decoherence on the XPM scheme is investigated in
detail. The application of our XPM scheme and the related discussion are
presented in Sec. \ref{Sec discussion}, while the conclusion is given in
Sec. \ref{Sec conclusion}

\begin{figure}[tbph]
\epsfig{file=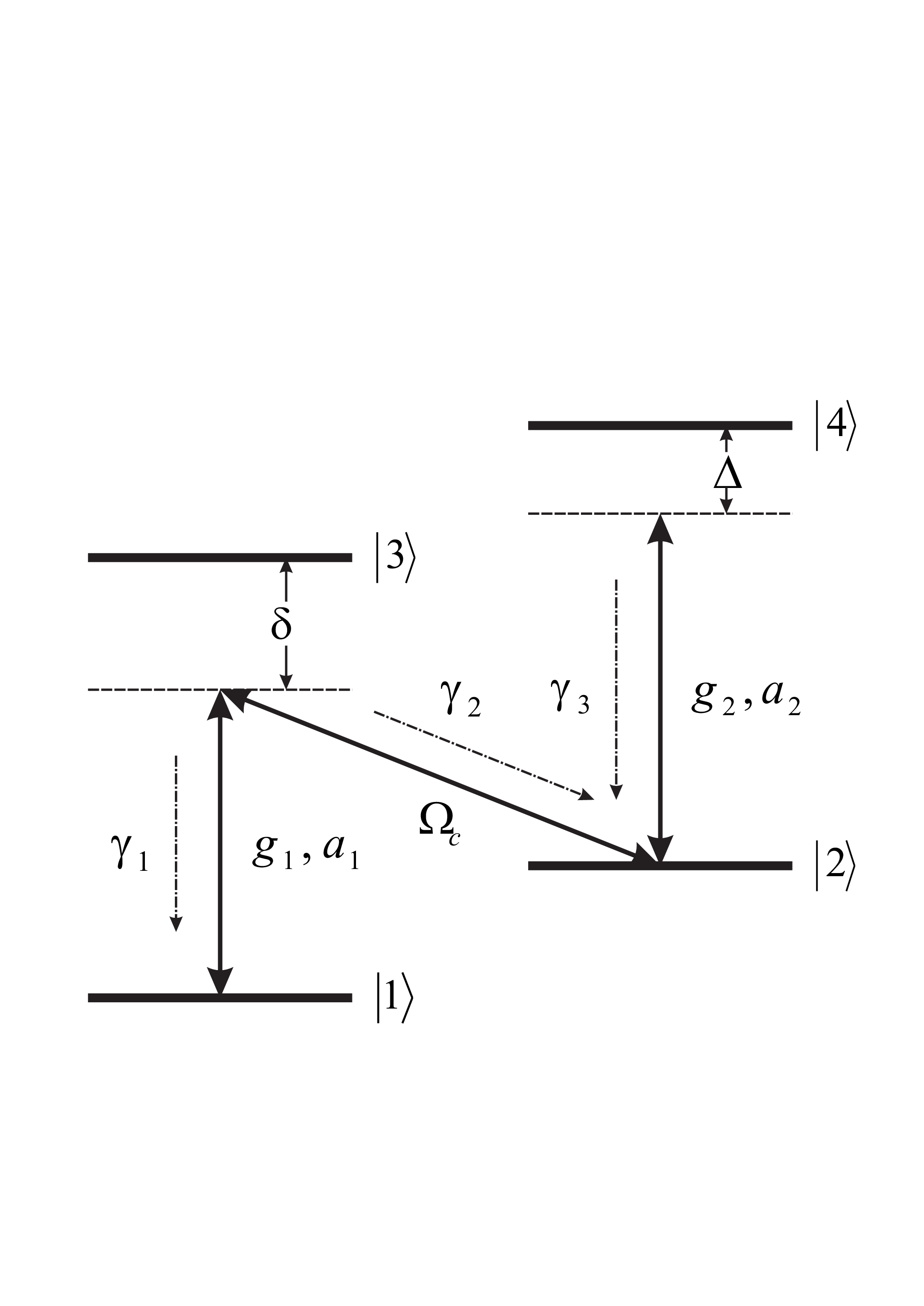, width=8cm}
\caption{(Color online). The N scheme for cross Kerr interaction. Two cavity
modes $a_{1}$ and $a_{2}$ induce the transitions $\left\vert 1\right\rangle
\rightarrow \left\vert 3\right\rangle $ and $\left\vert 2\right\rangle
\rightarrow \left\vert 4\right\rangle $ of the N type atom with coupling
strength $g_{1}$ and $g_{2}$ respectively. The $\protect\delta $ and $\Delta
$ are the corresponding detunings, while the factors $\protect\gamma _{1}$, $%
\protect\gamma _{2}$, and $\protect\gamma _{3}$ label the decay rates from
the upper levels to the lower levels. In addtion, an external pumping pulse
couples $\left\vert 2\right\rangle $ and $\left\vert 3\right\rangle $ with
Rabi frequency $\Omega _{c}$. The classical pumping is set to be in dark
resonance with the first cavity mode so that the states $\left\vert
1\right\rangle $, $\left\vert 2\right\rangle $ and $\left\vert
3\right\rangle $ exhibit electromagnetically induced transparency. }
\label{Fig FLM}
\end{figure}

\section{XPM induced by the N type atom\label{Sec theory}}

We start from the theoretical model of two cavity modes, labeled by their
annihilation operators $a_{1}$ and $a_{2}$, coupled to an atom with N type
level structure, as shown in Fig. \ref{Fig FLM}. The first mode couples the $%
\left\vert 1\right\rangle \leftrightarrow \left\vert 3\right\rangle $
transition with strength $g_{1}$ and detuning $\delta $, while the second
mode couples the $\left\vert 2\right\rangle \leftrightarrow \left\vert
4\right\rangle $ with strength $g_{2}$ and detuning $\Delta $. In addition,
a classical laser field is applied to drive the $\left\vert 2\right\rangle
\leftrightarrow \left\vert 3\right\rangle $ transition with strength $\Omega
_{c}$. The frequency of the classical field is set to be in Raman resonance
with the first mode. In the interaction picture, the Hamiltonian of the
system can be written as%
\begin{equation}
H_{sys}=H_{atom}+H_{int},  \label{Eq Hami}
\end{equation}%
with%
\begin{equation}
H_{atom}=\delta \sigma _{33}+\Delta \sigma _{44},
\end{equation}%
\begin{eqnarray}
H_{int} &=&i[g_{1}(a_{1}^{\dag }\sigma _{13}-\sigma
_{31}a_{1})+g_{2}(a_{2}^{\dag }\sigma _{24}-\sigma _{42}a_{2})
\label{Eq intheory} \\
&&+\Omega _{c}(\sigma _{23}-\sigma _{32})],  \notag
\end{eqnarray}%
where $\sigma _{jk}=\left\vert j\right\rangle \left\langle k\right\vert $
are the atomic raising and lowering operators for $j\neq k$ and population
operators for $j=k$. In the limit \cite{Adschmidt,Adwalls,Aderror}%
\begin{equation}
\left\vert \frac{g_{1}}{\Omega _{c}}\right\vert ^{2}\ll 1,\left\vert
g_{2}\right\vert \ll \left\vert \Delta \right\vert ,  \label{Eq criteria}
\end{equation}%
the atom evolves rapidly on the time scales relevant for the cavities. After
the adiabatical elimination of the atomic degrees of freedom, an effective
Kerr photon-photon interaction
\begin{equation}
H_{sys}\cong H_{eff}=-\frac{g_{2}^{2}}{\Delta }\frac{g_{1}^{2}}{\Omega
_{c}^{2}}a_{1}^{\dag }a_{1}a_{2}^{\dag }a_{2},  \label{Eq EFF}
\end{equation}%
can be obtained \cite{NScheme}.

The Kerr effect induced by the N type atoms has been investigated in the
past few years \cite%
{RebicPhDThesis,RebicGate,NScheme,NExp,NExTaiwan,NExShanxi}. The physics of $%
H_{eff}$ can be interpreted in an intuitive way: Suppose initially the atom
is in its ground state $\left\vert 1\right\rangle $. In the first step we
set $g_{2}=0$ so that $\left\vert 1\right\rangle $, $\left\vert
2\right\rangle $ and $\left\vert 3\right\rangle $ decouple from $\left\vert
4\right\rangle $. When the classical pumping is added, the states $%
\left\vert 1\right\rangle $ and $\left\vert 2\right\rangle $ form a dark
state, in which the destructive quantum interference between the $\left\vert
1\right\rangle \leftrightarrow \left\vert 3\right\rangle $ and $\left\vert
2\right\rangle \leftrightarrow \left\vert 3\right\rangle $ transition
branches cancels the dispersion and absorption \cite{ScullyQO} of the $a_{1}$
mode. The stationary population on the level $\left\vert 2\right\rangle $ is
thus determined by the photon number $a_{1}^{\dag }a_{1}$ as%
\begin{equation}
\left\langle \sigma _{22}\right\rangle \cong g_{1}^{2}a_{1}^{\dag
}a_{1}/\Omega _{c}^{2}.
\end{equation}%
When the second mode is added back to couple the $\left\vert 3\right\rangle
\leftrightarrow \left\vert 4\right\rangle $ transition dispersively, the
resulting AC Stark shift has the form $-g_{2}^{2}a_{2}^{\dag }a_{2}\sigma
_{22}/\Delta =-g_{1}^{2}g_{2}^{2}a_{1}^{\dag }a_{1}a_{2}^{\dag }a_{2}/\left(
\Delta \Omega _{c}^{2}\right) $.

Mathematically, to derive $H_{eff}$, we first write the evolution equations
of the operators $a_{1}$ and $a_{2}$ as
\begin{equation}
\frac{da_{1}}{dt}=g_{1}\sigma _{13}-\kappa _{1}a_{1}-\sqrt{2\kappa _{1}}%
a_{1in}(t),  \label{Eq a1}
\end{equation}%
\begin{equation}
\frac{da_{2}}{dt}=g_{2}\sigma _{24}-\kappa _{2}a_{2}-\sqrt{2\kappa _{2}}%
a_{2in}(t),  \label{Eq a2}
\end{equation}%
where $\kappa _{1,2}$ are the decay rates of the cavities, and $a_{1,2in}(t)$
are the input noise operators \cite{MilburnQO,ZollerQN}. Adiabatic
elimination of the atom allows us to express the atomic operators in terms
of the mode operators. Formally this is accomplished by setting the time
derivatives of all the $\sigma _{jk}$ to be zero. The stationary values $%
\left\langle \sigma _{jk}\right\rangle _{S}$ of the atomic operators $\sigma
_{jk}$ can be expanded as
\begin{equation}
\left\langle \sigma _{jk}\right\rangle _{S}\cong \sum\limits_{n=0}^{\infty
}\sigma _{jk}^{n}
\end{equation}%
where $\sigma _{jk}^{n}$ contains the multiplication of $n$ creation or
annihilation operators of the cavities. We further use $\sigma _{11}=1$ as
the starting value and iteratively determine the remaining coefficients of
the expansion. We thus get%
\begin{equation}
\left\langle \sigma _{13}\right\rangle _{S}=-\frac{g_{1}g_{2}^{2}}{i\Delta
\Omega _{c}^{2}}a_{2}^{\dag }a_{1}a_{2},  \label{Eq S13S}
\end{equation}%
\begin{equation}
\left\langle \sigma _{24}\right\rangle _{S}=-\frac{g_{1}^{2}g_{2}}{i\Delta
\Omega _{c}^{2}}a_{1}^{\dag }a_{1}a_{2}.  \label{Eq S24S}
\end{equation}%
Replacing $\sigma _{13}$ and $\sigma _{24}$ in Eqs. (\ref{Eq a1}) and (\ref%
{Eq a2}) by their stationary values in Eqs. (\ref{Eq S13S}) and (\ref{Eq
S24S}) we obtain%
\begin{equation}
\frac{da_{1}}{dt}=-\frac{g_{1}^{2}g_{2}^{2}}{i\Delta \Omega _{c}^{2}}%
a_{2}^{\dag }a_{1}a_{2}-\kappa _{1}a_{1}-\sqrt{2\kappa _{1}}a_{1in}(t),
\label{Eq A1}
\end{equation}%
\begin{equation}
\frac{da_{2}}{dt}=-\frac{g_{1}^{2}g_{2}^{2}}{i\Delta \Omega _{c}^{2}}%
a_{1}^{\dag }a_{1}a_{2}-\kappa _{2}a_{2}-\sqrt{2\kappa _{2}}a_{2in}(t).
\label{Eq A2}
\end{equation}%
The first terms in Eq. (\ref{Eq A1}) and (\ref{Eq A2}) yield an effective
Kerr interaction
\begin{equation}
H_{eff}=-\eta a_{2}^{\dag }a_{1}^{\dag }a_{1}a_{2},
\end{equation}%
where $\eta =g_{1}^{2}g_{2}^{2}/\left( \Delta \Omega _{c}^{2}\right) $ is
the Kerr coefficient.

The first order terms in the expansion of $\ \left\langle \sigma
_{13}\right\rangle _{S}$ and $\left\langle \sigma _{24}\right\rangle _{S}$
which represent the linear dispersion and absorption of the two cavities,
are vanishing. The suppression of linear susceptibilities is a result of the
quantum phase coherence in the atom \cite{ScullyQO}. When the decoherence of
the atom is taken into account, the phase coherence is broken and the linear
susceptibilities become nonzero. Also, the third order susceptibility $\eta $
is modified from real to complex. The derivation of Eqs. (\ref{Eq S13S}) and
(\ref{Eq S24S}) and the discussions about the influence of the atomic
decoherence are provided in the forthcoming sections and the Appendix.

\section{The N Scheme in circuit QED\label{Sec Natom}}

\begin{figure}[tbph]
\epsfig{file=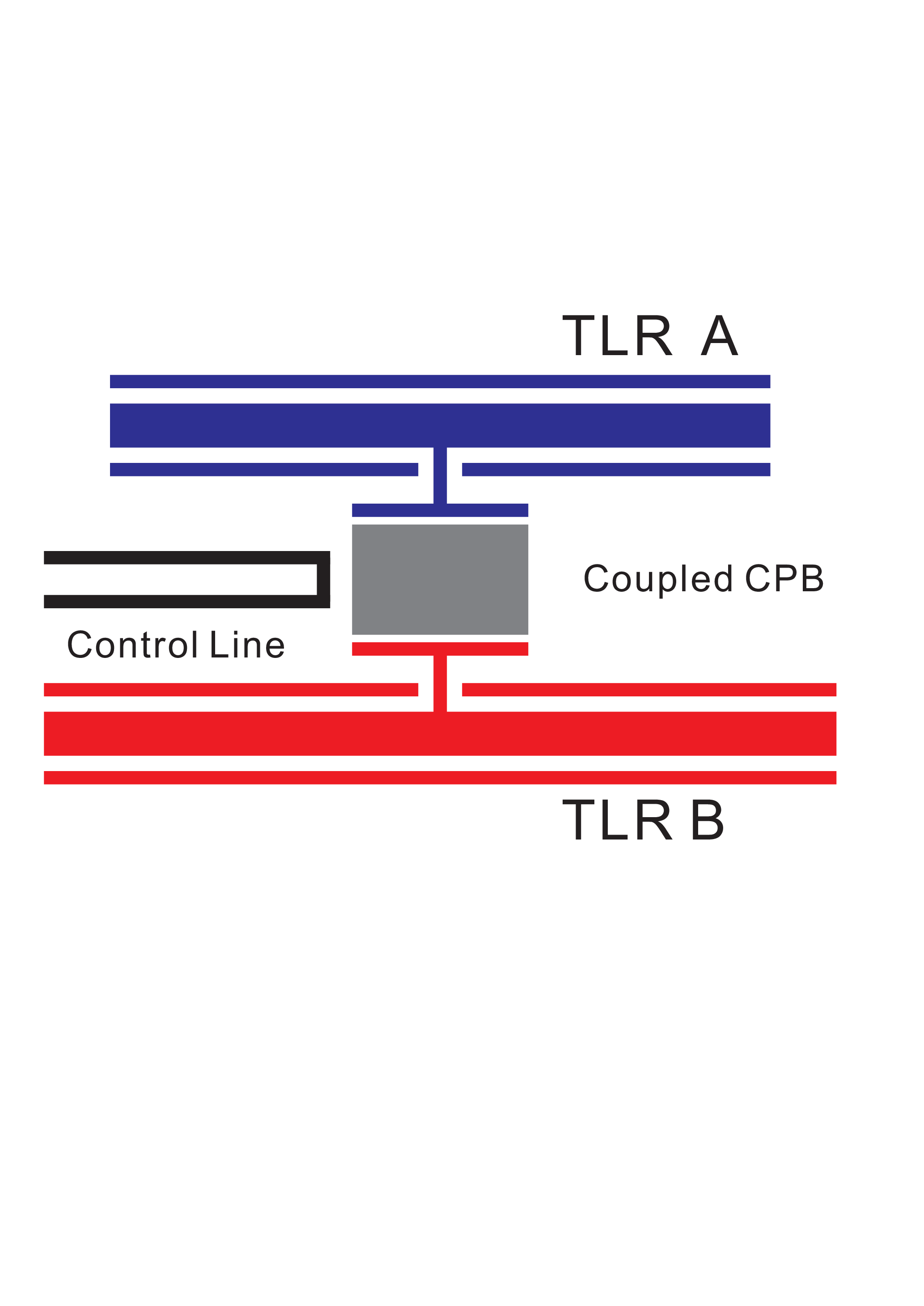, width=8cm}
\caption{(Color online). Schematic circuit for XPM between two TLRs. The TLR
A (blue) and the TLR B (red) are capacitively connected to a superconducting
molecule (gray) controlled by an external circuit (black). }
\label{Fig TLR}
\end{figure}

The N scheme described in Sec. \ref{Sec theory} can be realized in
superconducting quantum circuits. We consider a system of two TLRs
capacitively coupled to an artificial superconducting molecule, as shown in
Fig. \ref{Fig TLR}. In the following subsections we explicitly show that all
the required ingredients of the N level XPM proposal can be established in
the circuit QED system.

\subsection{The superconducting molecule: an artificial N type system}

\begin{figure}[tbph]
\epsfig{file=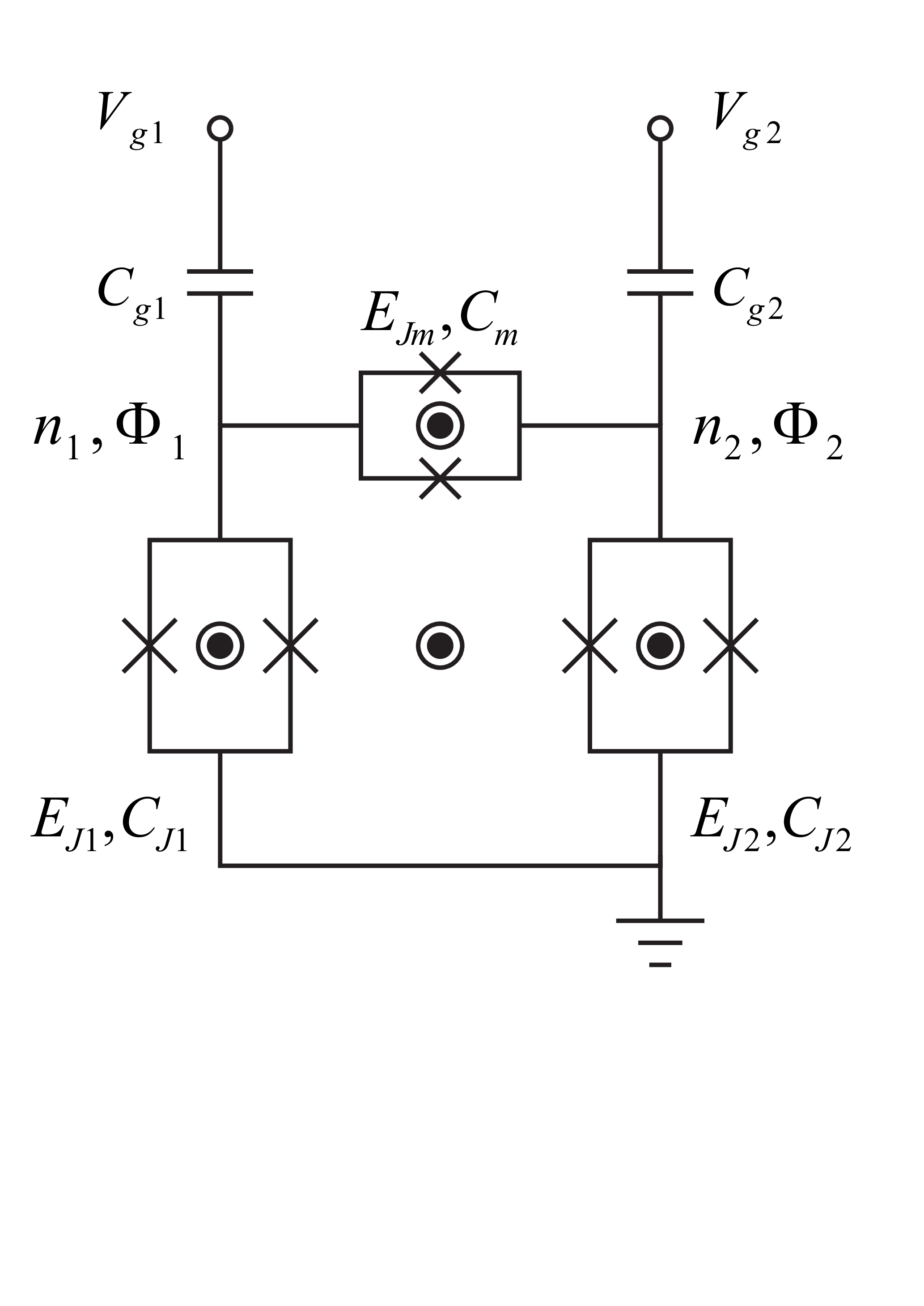, width=8cm}
\caption{(Color online). Schematic plot of the coupled CPBs. The crosses
label the Josephson junctions, while the round circle in the SQUID loops
represent the penetrating flux.}
\label{Fig CQ}
\end{figure}

The artificial molecule in Fig. \ref{Fig TLR} is constructed by two
superconducting Cooper pair boxes (CPB) \cite{NakamuraNature,Vion} coupled
by a SQUID \cite{TinkhamSC}, as sketched in Fig. \ref{Fig CQ}. Each CPB
consists of a small superconducting island connected to the ground electrode
by a symmetric SQUID with capacitance $C_{Ji}$ and tunable Josephson
coupling energy $E_{Ji}$ for $i=1,2$. The gate voltages $V_{gi}$ bias the
corresponding qubits via the gate capacitors $C_{gi}$. Finally, the CPBs are
connected to each other by a coupling SQUID with capacitance $C_{m}$ and
Josephson energy $E_{Jm}$. The Hamiltonian of the molecule reads%
\begin{eqnarray}
H_{\mathrm{0}} &=&4E_{m}(n_{1}-n_{g1})(n_{2}-n_{g2})-E_{Jm}\cos (\Phi
_{1}-\Phi _{2}) \\
&&+\sum\limits_{i=1,2}\left[ E_{ci}(n_{i}-n_{gi})^{2}-E_{Ji}\cos \Phi _{i}%
\right] ,  \notag
\end{eqnarray}%
where $n_{g1,2}=C_{g1,2}V_{g1,2}/2e$ denote the gate-induced charge numbers
on the CPBs, $\Phi _{1,2}$ are the canonical conjugate variables to $n_{1,2}$%
, $E_{c1,2}=2e^{2}C_{\Sigma 2,1}/(C_{\Sigma 1}C_{\Sigma 2}-C_{m}^{2})$ are
the effective Cooper-pair charging energies ($C_{\Sigma
i}=C_{gi}+C_{Ji}+C_{m}$ is the sum of all capacitances around the $i$th
qubit), and $E_{m}=e^{2}C_{m}/(C_{\Sigma 1}C_{\Sigma 2}-C_{m}^{2})$ is the
capacitive coupling strength between the CPBs. Near the co-degeneracy point $%
n_{g1}=n_{g2}=1/2$, we can use the two-level language%
\begin{equation}
n_{i}=(1+\sigma _{xi})/2,\cos \Phi _{i}=-\sigma _{zi}/2,
\end{equation}%
for $i=1,2$, to describe the molecule as%
\begin{eqnarray}
H_{\mathrm{0}} &=&E_{m}[\sigma _{x1}\sigma _{x2}-b_{0}(\sigma _{z1}\sigma
_{z2}+\sigma _{y1}\sigma _{y2})] \\
&&+\frac{1}{2}\sum\limits_{i=1,2}\left[ E_{bi}\sigma _{xi}+E_{Ji}\sigma _{zi}%
\right] ,  \notag
\end{eqnarray}%
where $E_{b1,2}=2\left[ E_{c1,2}\left( 1-2n_{g1,2}\right) +E_{m}\left(
1-2n_{g2,1}\right) \right] $ are the effective charge biases, and $%
b_{0}=E_{Jm}/4E_{m}$.

Being a Coulomb blockade device, the coupled-CPBs is very sensitive to noise
from charge degrees of freedom \cite{IthierPRB,SchrieflThesis}. By operating
the molecule at its optimal point, chosen such that the linear longitudinal
qubits-noise coupling vanishes, we can prolong its dephasing times by
several orders \cite{Vion}. For this reason, here we concentrate on the
behavior of the coupled CPBs at the co-degeneracy point $n_{g1}=n_{g2}=1/2$,
which can be verified as the optimal point later. Without loss of
generality, we further assume that the two CPBs are identical, i. e. $%
E_{J1}=E_{J2}=E_{J}$, $C_{J1}=C_{J2}=C_{J}$, and $C_{\Sigma 1}=C_{\Sigma
2}=C_{\Sigma }$. In this situation, the eigenstates and eigenvalues of $H_{%
\mathrm{0}}$ can be written as%
\begin{equation}
\begin{array}{c}
|1\rangle =-\sin \theta |00\rangle +\cos \theta |11\rangle , \\
|2\rangle =\left( -|01\rangle +|10\rangle \right) /\sqrt{2}, \\
|3\rangle =\left( |01\rangle +|10\rangle \right) /\sqrt{2}, \\
|4\rangle =\cos \theta |00\rangle +\sin \theta |11\rangle ,%
\end{array}
\label{Eq Level}
\end{equation}%
\begin{equation}
\begin{array}{c}
E_{1}=-E_{mN}-E_{m}b_{0}, \\
E_{2}=-E_{m}(1-2b_{0}), \\
E_{3}=E_{m}, \\
E_{4}=E_{mN}-E_{m}b_{0}.%
\end{array}%
\end{equation}%
where $E_{mN}=\sqrt{E_{J}^{2}+E_{m}^{2}(1-b_{0})^{2}}$, $E=\sqrt{%
E_{J}^{2}+E_{m}^{2}}$, and $\theta =[\arcsin (E_{m}/E)+\arcsin
(E_{J}b_{0}/2E_{mN})]/2$.

\begin{figure}[tbph]
\epsfig{file=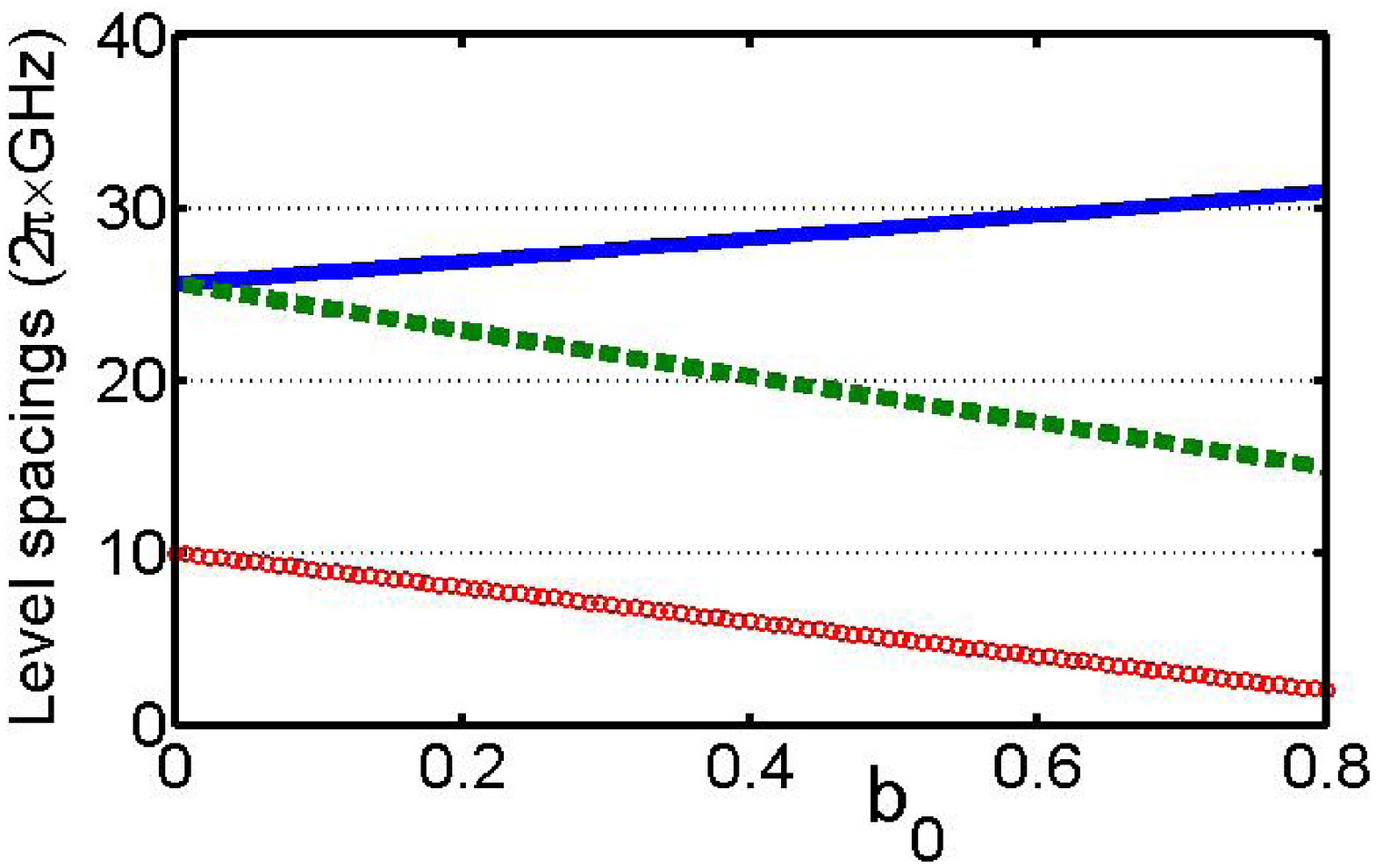, width=8cm}
\caption{(Color online). The level spacings versus $b_{0}$. The solid,
dashed, and round lines represent $E_{31}$, $E_{42}$, and $E_{32}$,
respectively. }
\label{Fig Level}
\end{figure}

The levels $\left\{ |1\rangle ,|2\rangle ,|3\rangle ,|4\rangle \right\} $ in
Eq. (\ref{Eq Level}) have the required N configuration in Sec. \ref{Sec
theory} and can be mapped to the four states in Fig. \ref{Fig FLM} one by
one. In experiments, $E_{J}$ is often larger than $E_{m}$, while the maximum
of $b_{0}$ is usually on the order of unity. We thus can modify the rough
energy scale of the molecule by tuning $E_{J}$. Moreover, $b_{0}$ can be
used to control the details of the energy structure. When $b_{0}=0$, the N
level is symmetric, i. e. $E_{4}-E_{2}=E_{3}-E_{1}$. Tuning $b_{0}$ to
non-zero value can break the symmetry. The influences of $b_{0}$ on the
level spacings $E_{42}=E_{4}-E_{2}$, $E_{31}=E_{3}-E_{1}$, and $%
E_{32}=E_{3}-E_{2}$ are shown in Fig. \ref{Fig Level}. In this calculation,
we set the parameters of the coupled CPBs as $E_{J}/2\pi =20$ \textrm{GHz }%
and $E_{m}/2\pi =5$ \textrm{GHz} \cite{Couple1,Couple2}. As plotted in Fig. %
\ref{Fig Level}, the level spacings go through very large range when $b_{0}$
varies in the region $b_{0}\in \left[ 0,0.8\right] $. Therefore, with the
tunability of $b_{0}$ and $E_{J}$, the level splits of the molecule can be
modulated at will. We further notice that the co-degeneracy bias point $%
n_{g1}=n_{g2}=1/2$ remains to be the optimal point for the molecule during
the tuning of $b_{0}$. The operators $\sigma _{x1}$ and $\sigma _{x2}$,
through which the system couples to the charge noise, have the form
\begin{equation}
\sigma _{x1}=\left[
\begin{array}{cccc}
0 & -\sin \phi & \cos \phi & 0 \\
-\sin \phi & 0 & 0 & \cos \phi \\
\cos \phi & 0 & 0 & \sin \phi \\
0 & \cos \phi & \sin \phi & 0%
\end{array}%
\right] ,
\end{equation}%
\begin{equation}
\sigma _{x2}=\left[
\begin{array}{cccc}
0 & \sin \phi & \cos \phi & 0 \\
\sin \phi & 0 & 0 & -\cos \phi \\
\cos \phi & 0 & 0 & \sin \phi \\
0 & -\cos \phi & \sin \phi & 0%
\end{array}%
\right] ,
\end{equation}%
where $\phi =\theta +\pi /4$. We thus verify that the diagonal entries of $%
\sigma _{x1}$ and $\sigma _{x2}$ are all zero, which indicates that the
linear longitudinal dephasing vanishes and the molecule is subject only to
the second order, quadratic dephasing.

Our design can be regarded as an innovation of the N type system proposed in
Ref. \cite{RebicSPM}, where Rebi\'{c} et. al have proposed to build an N
level molecule by two capacitively coupled CPBs. In that situation, the N
level structure is always symmetric on the optimal point. To get asymmetry,
additionally DC charge bias was needed and only moderate $\left\vert
E_{42}-E_{31}\right\vert $ can be obtained. Moreover, since the main
decoherence source of the charge based superconducting circuits is the
low-frequency charge noise, the DC bias in previous SPM scheme results
longitudinal dephasing which severely damages the phase coherence of the
molecule. Therefore, compared with previous design, our introduction of the
coupling SQUID can offer more tunability and robustness to the artificial
molecule.

\subsection{The classical pulse}

We proceed to show how to implement the classical pumping connecting $%
|2\rangle $ and $|3\rangle $. Penetrating the external flux into the SQUID
loops of the CPBs can couple microwave pulses with the molecule through the
operators $\sigma _{z1}$ and $\sigma _{z2}$, which have the matrix form

\begin{equation}
\sigma _{z1}=\left[
\begin{array}{cccc}
-\cos 2\theta & 0 & 0 & -\sin 2\theta \\
0 & 0 & -1 & 0 \\
0 & -1 & 0 & 0 \\
-\sin 2\theta & 0 & 0 & \cos 2\theta%
\end{array}%
\right] ,
\end{equation}%
\begin{equation}
\sigma _{z2}=\left[
\begin{array}{cccc}
-\cos 2\theta & 0 & 0 & -\sin 2\theta \\
0 & 0 & 1 & 0 \\
0 & 1 & 0 & 0 \\
-\sin 2\theta & 0 & 0 & \cos 2\theta%
\end{array}%
\right] .
\end{equation}%
The matrix elements of $\sigma _{z1}$ and $\sigma _{z2}$ support the
transitions inside the Hilbert subspaces $\mathrm{span}\left\{ |1\rangle
,|4\rangle \right\} $ and $\mathrm{span}\left\{ |2\rangle ,|3\rangle
\right\} $. Therefore, we can apply an AC flux pulse to the SQUID loop of
the first qubit to modulate $E_{J1}$ with amplitude $\Omega _{Ex}$ and
frequency $\omega _{p}$. When $\omega _{p}$ is tuned close to $E_{32}$, the $%
|2\rangle \leftrightarrow |3\rangle $ transition is effectively induced
while the $|1\rangle \leftrightarrow |4\rangle $ transition can be neglected
due to frequency mismatch. To further suppress the unwanted $|1\rangle
\leftrightarrow |4\rangle $ transition, we can synchronizely modulate $%
E_{J2} $ with the same amplitude and opposite phase as that of $E_{J1}$. In
this way we get a pumping Hamiltonian
\begin{eqnarray}
H_{pump} &=&\Omega _{Ex}\cos \omega _{P}t(\sigma _{z1}-\sigma _{z2})
\label{Eq pump} \\
&\cong &-\Omega _{Ex}\left[ \exp \left( i\omega _{P}t\right) |2\rangle
\left\langle 3\right\vert +\exp \left( -i\omega _{P}t\right) |3\rangle
\left\langle 2\right\vert \right]  \notag
\end{eqnarray}%
which establishes the pure $|2\rangle \leftrightarrow |3\rangle $\
transition.

\subsection{Coupling between the TLRs and the molecule}

We consider the capacitive coupling between the TLRs and the molecule. As
shown in Fig. \ref{Fig TLR}, The TLR A (blue) has length $L_{A}$ and
fullwave frequency $\omega _{A}$, while the TLR B (red) has length $L_{B}$
and fullwave frequency $\omega _{B}$. The Hamiltonian of the two individual
TLRs are
\begin{equation}
H_{Cavity}=\omega _{A}a^{\dag }a+\omega _{B}b^{\dag }b,
\end{equation}%
where $\omega _{A,B}=2\pi /\left( L_{A,B}\sqrt{Fc}\right) $ are the
eigenfrequencies, $F$ and $c$ being the inductance and capacitance per unit
length, $L_{A,B}$ being the length of the two TLRs, and $a$, $b$ are the
annihilation operators of the fullwave modes, respectively. The quantized
voltages of the TLRs are
\begin{eqnarray}
V_{A}(x) &=&\sqrt{\omega _{A}/L_{A}c}(a^{\dag }+a)\cos (2\pi x/L_{A}),x\in %
\left[ -L_{A}/2,L_{A}/2\right] , \\
V_{B}(x) &=&\sqrt{\omega _{B}/L_{B}c}(b^{\dag }+b)\cos (2\pi x/L_{B}),x\in %
\left[ -L_{B}/2,L_{B}/2\right] .  \notag
\end{eqnarray}

When the two TLRs are capacitively connected with the molecule, the
interaction Hamiltonian has the general form%
\begin{equation}
H_{cc}=V_{1}\left[ C_{A1}V_{A}(x_{A1})+C_{B1}V_{B}(x_{B1})\right] +V_{2}%
\left[ C_{A2}V_{A}(x_{A2})+C_{B2}V_{B}(x_{B2})\right] ,
\end{equation}%
where the $C_{Aj}$, $C_{Bj}$ for $j=1,2$ are the coupling capacitance
between the $j$th qubit and the TLRs, the $x_{Aj}$, $x_{Bj}$ for $j=1,2$ are
the locations of the coupling capacitor, and $V_{1,2}=$ $\left[ C_{\Sigma
}\sigma _{x1,2}+C_{m}\sigma _{x2,1}\right] /(C_{\Sigma }^{2}-C_{m}^{2})$ are
the quantized voltages of the first and second CPB. We further expand $%
H_{cc} $ as
\begin{eqnarray}
H_{cc} &=&h_{A1}(a^{\dag }+a)\sigma _{x1}+h_{A2}(a^{\dag }+a)\sigma _{x2} \\
&&+h_{B1}(b^{\dag }+b)\sigma _{x1}+h_{B2}(b^{\dag }+b)\sigma _{x2}  \notag
\end{eqnarray}%
where $h_{A,B;1,2}$ are the coupling factors which can be written as%
\begin{eqnarray}
h_{A,B;1,2} &=&\sqrt{\frac{\omega _{A,B}}{L_{A,B}c}}\frac{e}{\left(
C_{\Sigma }^{2}-C_{m}^{2}\right) }[C_{A,B;1,2}\cos (\frac{2\pi x_{A,B;1,2}}{%
L_{A,B}})C_{\Sigma 2,1} \\
&&+C_{A,B;2,1}\cos (\frac{2\pi x_{A,B;2,1}}{L_{A,B}})C_{m}].  \notag
\end{eqnarray}

The operators $\sigma _{x1}$ and $\sigma _{x2}$ have matrix elements which
induce the transitions between subspace $\mathrm{span}\left\{ |1\rangle
,|4\rangle \right\} $ and $\mathrm{span}\left\{ |2\rangle ,|3\rangle
\right\} $. We can thus tune the $E_{J}$ and $b_{0}$\ so that $E_{42}$ and $%
E_{31}$ are close to the cavity mode frequencies $\omega _{A,B}$. In this
case the TLRs can effectively only induce the $|1\rangle \leftrightarrow
|3\rangle $ and $|2\rangle \leftrightarrow |4\rangle $ transitions. The
remaining $|1\rangle \leftrightarrow |2\rangle $ and $|3\rangle
\leftrightarrow |4\rangle $ transitions are suppressed due to frequency
selection. With the rotating wave approximation, $H_{cc}$ finally reads
\begin{eqnarray}
H_{cc} &=&g_{A1}(\sigma _{13}a^{\dag }+\sigma _{31}a)+g_{B1}(\sigma
_{13}b^{\dag }+\sigma _{31}b)  \label{Eq cc} \\
&&+g_{A2}(\sigma _{24}a^{\dag }+\sigma _{42}a)+g_{B2}(\sigma _{24}b^{\dag
}+\sigma _{42}b),  \notag
\end{eqnarray}%
where the coupling factors $g_{A,B;1,2}$ are%
\begin{eqnarray}
g_{A,B;1} &=&\cos \varphi \left( h_{A,B;1}+h_{A,B;2}\right) , \\
g_{A,B;2} &=&\cos \varphi \left( h_{A,B;1}-h_{A,B;2}\right) .  \notag
\end{eqnarray}

\subsection{XPM in circuit QED}

$H_{cc}$ in Eq. (\ref{Eq cc}) has a more general form than the atom-photon
coupling terms in Eq. (\ref{Eq intheory}). The flexibility of
superconducting devices allow us to choose suitable values and locations of
the coupling capacitors in order to get desired coupling configurations. The
setting of the locations of the capacitors does not require to put the CPBs
in distant places, because the TLRs can be fabricated in a zig-zag form. A
trivial case is that $C_{B1}=C_{B2}=0$, i. e. the molecule is connected only
to the TLR A, which results the SPM of circuit QED \cite{RebicSPM}. To
establish the XPM described in Sec. \ref{Sec theory}, we set the coupling
capacitors to have capacitances $C_{B1}=C_{B2}$ and $C_{A1}=C_{A2}$. In
addition, the location of the capacitors are selected as $x_{A1}=x_{A2}=0$
and $x_{B1}=L_{B}/2-x_{B2}=L_{B}/8$. In this way, $h_{A1}=h_{A2}$, $%
h_{B1}=-h_{B2}$ and the resulting $H_{cc}$ is given by%
\begin{equation}
H_{cc}=g_{A1}(\sigma _{13}a^{\dag }+\sigma _{31}a)+g_{B2}(\sigma
_{24}b^{\dag }+\sigma _{42}b).  \label{Eq int}
\end{equation}%
Up to a trivial $i$ factor, $H_{cc}$ in Eq. (\ref{Eq int}) has the\ same
atom-photon coupling form as that of $H_{int}$ in Eq. (\ref{Eq intheory}),
i. e. the TLRs A and B replace the role of the cavities $a_{1}$ and $a_{2}$,
respectively.

Combining Eqs. (\ref{Eq Level}), (\ref{Eq pump}), and (\ref{Eq int}), We
develop all the required elements of the N level XPM scheme in the circuit
QED system. When we tune the classical pumping to be in dark resonance with
the TLR A, an effective Kerr interaction hamiltonian%
\begin{equation}
H_{eff}=-\frac{g_{B2}^{2}}{\Delta }\frac{g_{A1}^{2}}{\Omega _{Ex}^{2}}%
a^{\dag }b^{\dag }ab,
\end{equation}%
can be induced between the two TLRs.

The Kerr coefficient $\chi _{3}=g_{A1}^{2}g_{B2}^{2}/\left( \Delta \Omega
_{Ex}^{2}\right) $ can be estimated based on the reported experiments. Since
the coupling strength between a TLR and a qubit as large as $300$ \textrm{MHz%
} has been achieved, we set $g_{1}/2\pi =g_{2}/2\pi =300$ GHz. In addition,
the classical pumping strength $\Omega _{Ex}$ and the detuning $\Delta
=E_{42}-\omega _{B}$ can be chosen as $\Omega _{Ex}/2\pi =1.5$ GHz and $%
\Delta /2\pi =1.5$ \textrm{GHz} to fulfill the adiabatical condition in Eq. (%
\ref{Eq criteria}). We then get $\chi _{3}/2\pi \cong 2.5$ \textrm{MHz},
which has already exceeded the observed Kerr strength obtained by exploiting
a large Josephson junction connected with the TLR in the recent papers \cite%
{Intermode,DiVincenzo,VionNonlinear}.

\section{The role of atomic decoherence in XPM\label{Sec deco},}

We discuss the influence of molecule's decoherence on the XPM. As shown in
Fig. \ref{Fig FLM}, in the first step we consider only the decay processes $%
|3\rangle \rightarrow |1\rangle $, $|3\rangle \rightarrow |2\rangle $, and $%
|4\rangle \rightarrow |2\rangle $, with decay rates $\gamma _{1}$, $\gamma
_{2}$, and $\gamma _{3}$, respectively. In this case the polarizations $%
\left\langle \sigma _{13}\right\rangle _{S}$ and $\left\langle \sigma
_{24}\right\rangle _{S}$ have the form
\begin{eqnarray}
\left\langle \sigma _{13}\right\rangle _{S} &=&-\frac{g_{1}g_{2}^{2}}{\left(
\gamma _{3}+i\Delta \right) \Omega _{c}^{2}}a_{2}^{\dag }a_{1}a_{2}, \\
\left\langle \sigma _{24}\right\rangle _{S} &=&-\frac{g_{1}^{2}g_{2}}{\left(
\gamma _{3}+i\Delta \right) \Omega _{c}^{2}}a_{1}^{\dag }a_{1}a_{2},  \notag
\end{eqnarray}%
which results a complex Kerr coefficient%
\begin{equation}
\chi _{3}=\frac{g_{1}^{2}g_{2}^{2}}{\left( -i\gamma _{3}+\Delta \right)
\Omega _{c}^{2}}.
\end{equation}%
\ We notice that $\chi _{3}$ does not depend on either $\gamma _{1}$ or $%
\gamma _{2}$. In addition, the first order terms of $\left\langle \sigma
_{13}\right\rangle _{S}$ and $\left\langle \sigma _{24}\right\rangle _{S}$
are still zero. These two effect can be explained by the EIT in the lowest
levels of the molecule. The destructive interference between the $|3\rangle
\leftrightarrow |1\rangle $ and $|3\rangle \leftrightarrow |2\rangle $
transition branches cancels the first order terms of $\left\langle \sigma
_{13}\right\rangle _{S}$ and $\left\langle \sigma _{24}\right\rangle _{S}$
and thus the linear susceptibilities of the TLRs. Moreover, the coherent
population trapping significantly suppress the population $\left\langle
\sigma _{33}\right\rangle _{S}$. As a result, the decay channels $|3\rangle
\leftrightarrow |1\rangle $ and $|3\rangle \leftrightarrow |2\rangle $
become irrelevant. Therefore, we use the relative decay rate $R$ define by $%
R=\left\vert \Im (\chi _{3})/\Re (\chi _{3})\right\vert $ to characterize
the molecule-induced cavity decay. Since $\gamma _{3}/2\pi $ has been pushed
to the order of $0.5$ \textrm{MHz} in recent experiments \cite{Transmon},
while $\Delta $ can be usually set to be on the order of \textrm{GHz}, $%
R=\gamma _{3}/\Delta $ is estimated to be in the range $\left[
10^{-3},10^{-4}\right] $.

\begin{figure}[tbph]
\epsfig{file=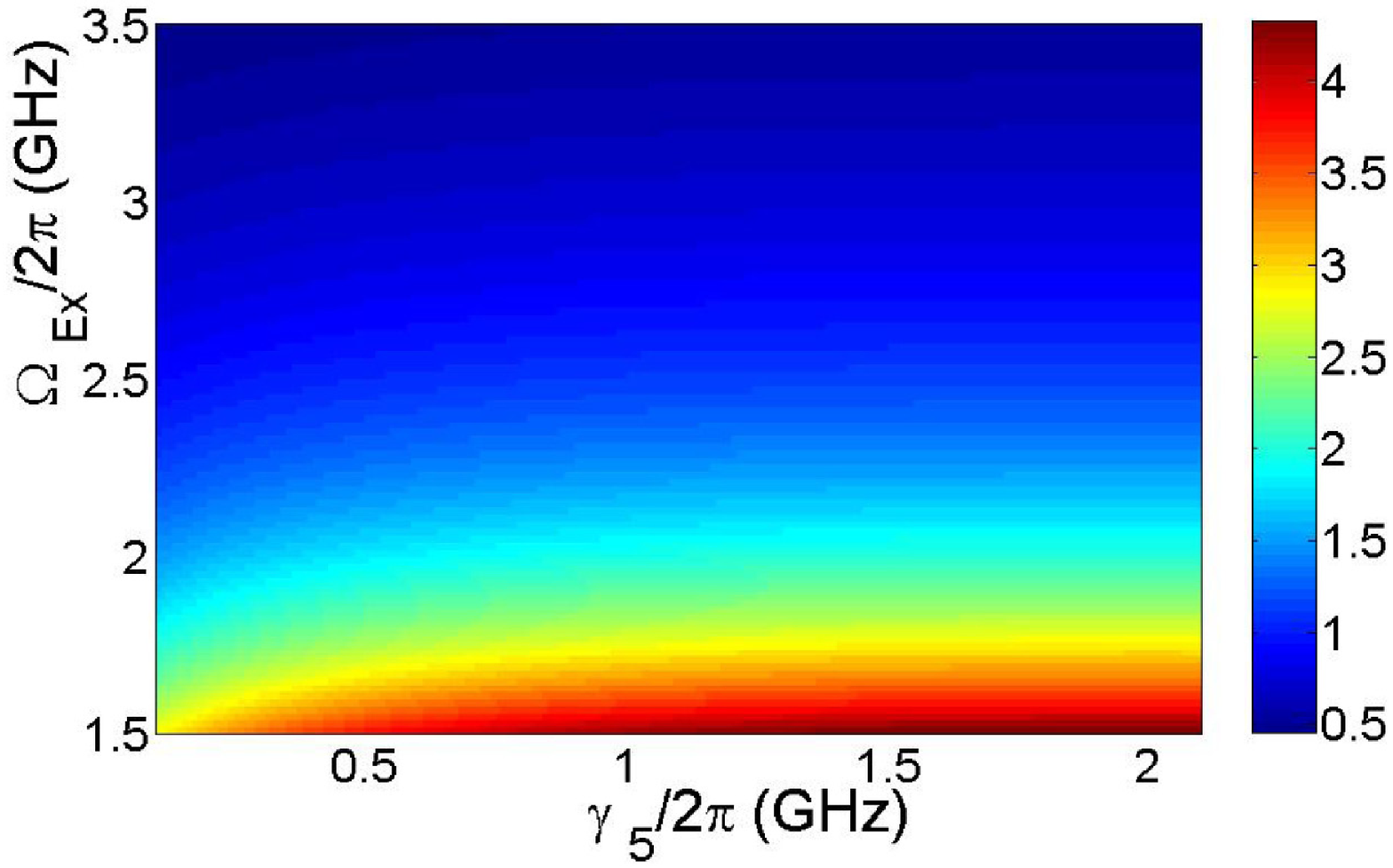, width=8cm}
\caption{(Color online). The Kerr nonlinearity $\protect\chi _{3}^{\prime
}=\Re(\protect\chi _{3})$ versus $\Omega _{Ex}$ and $\protect\gamma _{5}$.}
\label{Fig kappa3}
\end{figure}

\begin{figure}[tbph]
\epsfig{file=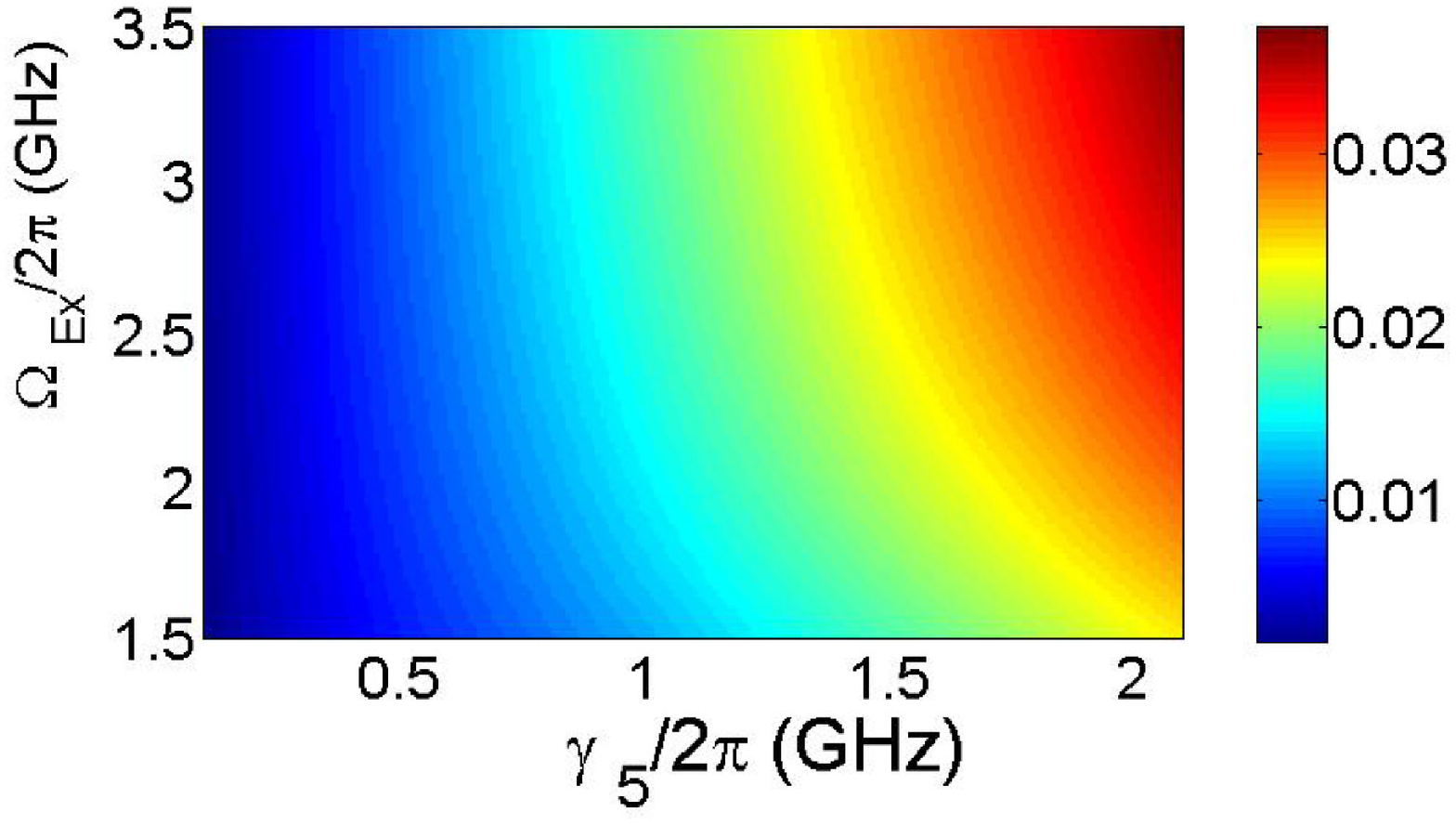, width=8cm}
\caption{(Color online). The linear dispersion factor $\protect\chi %
_{1}^{\prime }/\protect\chi _{3}^{\prime }$ versus $\Omega _{Ex}$ and $%
\protect\gamma _{5}$. }
\label{Fig rir}
\end{figure}

\begin{figure}[tbph]
\epsfig{file=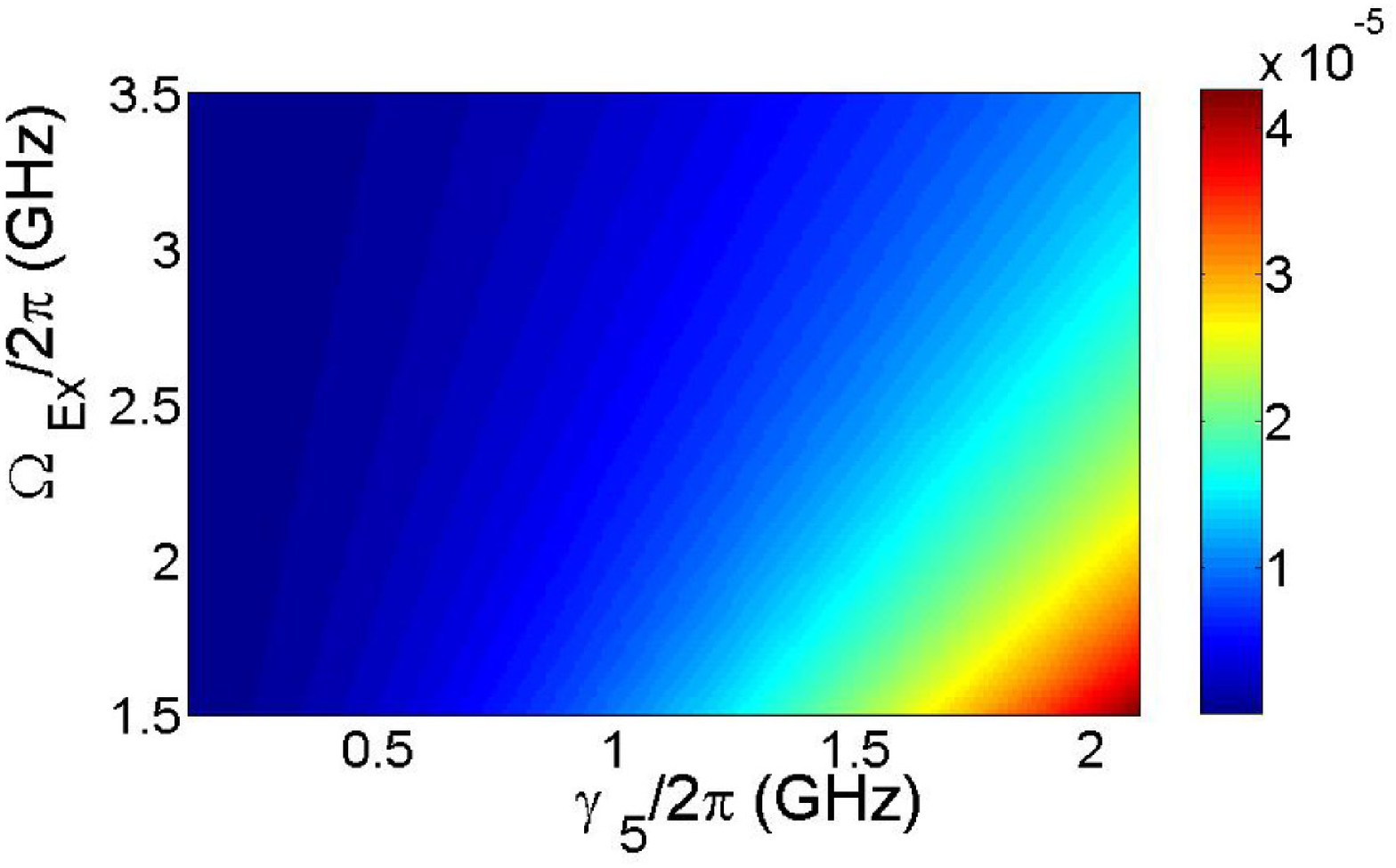, width=8cm}
\caption{(Color online). The linear absorption factor $\protect\chi %
_{1}^{\prime \prime }/\protect\chi _{3}^{\prime }$ versus $\Omega _{Ex}$ and
$\protect\gamma _{5}$.}
\label{Fig rim}
\end{figure}

\begin{figure}[tbph]
\epsfig{file=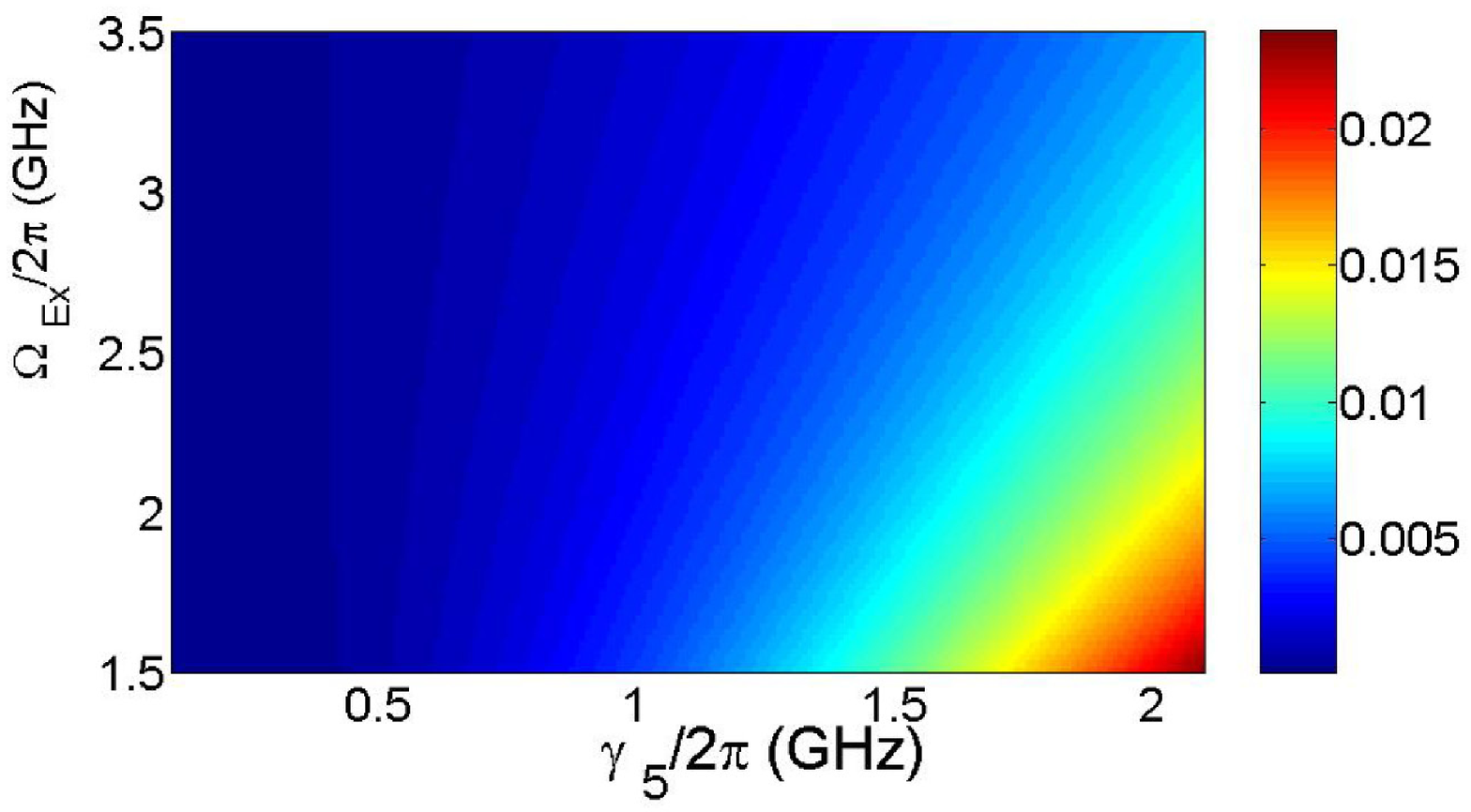, width=8cm}
\caption{(Color online). The nonlinear absorption factor $\Im(\protect\chi %
_{3})/\protect\chi _{3}^{\prime }$ versus $\Omega _{Ex}$ and $\protect\gamma %
_{5}$.}
\label{Fig r3m}
\end{figure}

We further take the decay $|2\rangle \rightarrow |1\rangle $ and dephasing
between $|2\rangle $ and $|1\rangle $ into account. The corresponding
decoherence rates are $\gamma _{4}$ and $\gamma _{\phi }$. These two
processes can break the phase coherence between $|2\rangle $ and $|1\rangle $%
, which is essential in the previous suppression of linear dispersion and
absorption. In this circumstance the first order of $\left\langle \sigma
_{13}\right\rangle _{S}$ emerges. It takes the form
\begin{equation}
\sigma _{13}^{1}=-\gamma _{5}g_{A1}a/B,
\end{equation}%
where $\gamma _{5}=\gamma _{4}+\gamma _{\phi }$ and $B=\Omega
_{Ex}^{2}+\gamma _{1}\gamma _{5}+\gamma _{2}\gamma _{5}+i\delta \gamma _{5}$%
. The resulting linear susceptibility is
\begin{equation}
\chi _{1}=g_{A1}^{2}\gamma _{5}/B.
\end{equation}%
the real part $\chi _{1}^{\prime }$ of $\chi _{1}$ which describes the
decoherence induced dispersion of the TLR A is approximately $%
g_{A1}^{2}\gamma _{5}/\Omega _{Ex}^{2}$. With $\gamma _{5}/2\pi \cong 0.5$ $%
\mathrm{MHz}$, $\chi _{1}^{^{\prime }}/2\pi $ is lower than $0.01$ \textrm{%
MHz}. The imaginary part $\chi _{1}^{\prime \prime }$ of $\chi _{1}$, which
describes the decoherence induced dissipation of the TLR A, can be
approximated as $\chi _{1}^{\prime \prime }\cong g_{A1}^{2}\gamma _{5}\delta
\gamma _{5}/\Omega _{Ex}^{4}$. It is even smaller than $\chi _{1}^{\prime }$
by several orders. The analytic form of $\chi _{3}$ in the presence of the
decoherence $|2\rangle \rightarrow |1\rangle $ and the dephasing becomes
complicated. We thus numerically calculate the Kerr coefficient $\chi
_{3}^{\prime }=\Re (\chi _{3})$ versus the pumping $\Omega _{Ex}$ and the
decoherence rate $\gamma _{5}$. The result is shown in Fig. \ref{Fig kappa3}%
. We find that $\chi _{3}^{\prime }$ depends weakly on $\gamma _{5}$. When $%
\gamma _{5}$ becomes non zero, $\chi _{3}^{\prime }/2\pi $ remains to be
larger than $1$ \textrm{MHz}. We further calculate the dispersion factor $%
\chi _{1}^{\prime }/\chi _{3}^{\prime }$, the linear absorption factor $\chi
_{1}^{\prime \prime }/\chi _{3}^{^{\prime }}$, and the nonlinear absorption
factor $\Im (\chi _{3})/\chi _{3}^{\prime }$ and plot the results in Figs. %
\ref{Fig rir}, \ref{Fig rim}, and \ref{Fig r3m}. We see that the unwanted
effects increase rapidly with increasing $\gamma _{5}$. Nevertheless, in the
region of current technology $\gamma _{5}/2\pi \cong 0.5$ $\mathrm{MHz}$,
these effects are still smaller than the XPM\ by at lease two orders of
magnitude. If the decoherence effects can be suppressed to the range $\gamma
_{5}/2\pi \cong 0.1$ $\mathrm{MHz}$ in the future experiments, a more pure
XPM with negligible dispersion and absorption can be achieved.

\section{Applications and Discussion\label{Sec discussion}}

As mentioned in the previous section, the molecule induced fluctuation of
the TLRs is approximately two to three orders lower than the cross Kerr
nonlinearity which is on the level \textrm{MHz}. In addition, with current
technology, the Q factor of a single TLR has already been pushed to the
order of $10^{6}$ \cite{FrunzioIEEE}, which yields the decay rate on the
level $10$ \textrm{kHz}. Since both the intrinsic decoherence of the TLRs
and the molecule induced decoherence are all very small compared with the
strength of Kerr nonlinearity, several important inter-TLR quantum
operations in which only few photons are involved in each TLR could be
realized with very high fidelities. Recently there are proposals of using
the TLRs as qubits to realize scalable quantum computing \cite%
{Mariantoni,DuTLR}. The TLR states with zero and one photon are used as the
logical $\left\vert 0\right\rangle $ and $\left\vert 1\right\rangle $.
Therefore, our proposed Kerr nonlinearity can obviously be exploited to
establish the control phase gates in this system. Another application of the
cross Kerr nonlinearity is the generation of the inter-TLR macroscopic
maximal entangled cat state \cite{Sanders,LukinPRL,Reventangle}. In the
first step we tune the molecule decoupled with the TLRs. We then use
classical pulses to pump the TLRs to the factorized coherent state $%
\left\vert \alpha \right\rangle _{A}\left\vert \beta \right\rangle _{B}$ and
then adiabatically tune the molecule to establish XPM with strength $\chi
_{3}$ between the TLRs. After a time $t=\pi /\chi _{3}$, the initial $%
\left\vert \alpha \right\rangle _{A}\left\vert \beta \right\rangle _{B}$
evolves to a state
\begin{equation}
\left\vert \Psi \right\rangle =\frac{1}{2}(\left\vert \alpha \right\rangle
_{A}\left\vert \beta \right\rangle _{B}+\left\vert -\alpha \right\rangle
_{A}\left\vert \beta \right\rangle _{B}+\left\vert \alpha \right\rangle
_{A}\left\vert -\beta \right\rangle _{B}-\left\vert -\alpha \right\rangle
_{A}\left\vert -\beta \right\rangle _{B})
\end{equation}%
Although the coherent states are not orthogonal, the overlap $\left\vert
\left\langle \beta |\alpha \right\rangle \right\vert $ is only $10^{-7}$ for
a moderate $\left\vert \beta -\alpha \right\vert =4$. Superpositions of this
kind have no classical counterpart and correspond to Schr\"{o}dinger catlike
states. Therefore, the state $\left\vert \Psi \right\rangle $ can be viewed
as the macroscopic entangled cat state, which can be used as logical
elements in quantum computation and can offer further insight into the
boundary of classical-quantum worlds. We mention that since the coherent
state spreads over the whole Fock space, such an entangled cat state can
hardly be prepared by the generalized Eberly-Law algorithm, which has been
used to generate the NOON states in the circuit QED system \cite%
{SimmondsTLR,MartinisTLR}.

An alternative coupling configuration we can achieve is that the TLR A
couples the $|1\rangle \leftrightarrow |3\rangle $ transition while the TLR
B couples both the $|1\rangle \leftrightarrow |3\rangle $ and $|2\rangle
\leftrightarrow |4\rangle $ transitions. The coupling between the TLR B and
the $|1\rangle \leftrightarrow |3\rangle $ transition\ can result from
either the intention of the experiment or the fabrication errors of the
coupling capacitors. In this situation, the difference between the mode
frequencies plays an important role. When $\omega _{A}=\omega _{B}$, an
effective Hamiltonian%
\begin{equation}
H_{eff}=-\frac{g_{B2}^{2}}{\Delta }\frac{g_{A1}^{2}}{\Omega _{Ex}^{2}}%
a^{\dag }b^{\dag }ab-\frac{g_{B2}^{2}}{\Delta }\frac{g_{B1}^{2}}{\Omega
_{Ex}^{2}}b^{\dag }b^{\dag }bb,
\end{equation}%
in which the self Kerr effect and the cross Kerr effect co-exist is
obtained. When $\omega _{A}\neq \omega _{B}$, since the TLR B is not
resonant with the classical pumping field, it can hardly transfer population
to the state $|2\rangle $, the effective TLR-TLR Hamiltonian thus remains
the form $-g_{A1}^{2}g_{B2}^{2}a^{\dag }b^{\dag }ab/(\Delta \Omega
_{Ex}^{2}) $. From this point of view, it is favorable to use TLRs with
different eigenfrequencies to realize the XPM, since the difference of their
frequencies makes the scheme more robust against the fabrication errors.

Recently theoretical and experimental works have suggested that the large
Josephson junction can be used to produce SPM and XPM for the TLRs \cite%
{Intermode,DiVincenzo,VionNonlinear,DuTLR}. By connecting a large Josephson
junction with the TLRs, one can obtain the Kerr nonlinearity from the Taylor
expansion of the Josephson energy up to the fourth order. The Kerr
nonlinearity obtained in this way is smaller than what we have proposed,
often on the order of hundreds of \textrm{kHz}. Moreover, the critical
current noise and the flux noise in the large Josephson junction results
large fluctuations of the Josephson energy and consequently the severe
linear dispersion of the TLRs. To eliminate the effect of the fluctuation,
one have to use additional spin-echo like technique \cite{DuTLR} which
complicate the quantum gate sequence. Compared with these schemes, our
alternative XPM scheme may be more robust against the noises and may offer
more pure Kerr nonlinearity with suppressed frequency drifts.

\section{Conclusion\label{Sec conclusion}}

In conclusion, in this paper we have shown that an \textquotedblleft
artificial\textquotedblright\ multilevel system in circuit QED produces the
effective XPM between two TLRs with strength much larger than previously
known. The obtained XPM is very robust against the noises in solid state
system. Compared with the XPM strength, the accompanying dispersion and
absorption is negligible. Various QIPs can be implemented in this
architecture. This work may offer improvement to the future scalable quantum
computation in superconducting devices.

\begin{acknowledgments}
We thank Prof. Y. F. Zhu, S. Rebi\'{c}, Z. W. Zhou, and Y. F. Xiao for
fruitful discussions. This work is funded by the start funding of HUST, No.
01-24-012018 and No. 01-24-012030, the Natural Science Foundation of Hubei
province, and the National Natural Science Foundation of China, Grant No.
60871018.
\end{acknowledgments}

\appendix

\section{The effective XPM Hamiltonian}

The derivation of Eqs. (\ref{Eq S13S}) and (\ref{Eq S24S}) is provided in
this Appendix. As shown in Fig. \ref{Fig FLM}, taking the decoherence
processes $\left\vert 3\right\rangle \rightarrow \left\vert 1\right\rangle $%
, $\left\vert 3\right\rangle \rightarrow \left\vert 2\right\rangle $, and $%
\left\vert 4\right\rangle \rightarrow \left\vert 2\right\rangle $ of the
atom and the decay of the two cavities into account, we write the
system-bath Hamiltonian as
\begin{equation}
H_{whole}=H_{sys}+H_{bath},
\end{equation}%
where $H_{damp}$ represents the coupling of the system to reservoir
mediating cavity decay and spontaneous emission; it takes the form \cite%
{ZollerQN}%
\begin{eqnarray}
H_{damp} &=&\sum_{j=1,2}\int\limits_{-\infty }^{+\infty }\omega b_{j}^{\dag
}(\omega )b_{j}(\omega )d\omega +\sum_{j=1,2,3}\int\limits_{-\infty
}^{+\infty }\omega \beta _{j}^{\dag }(\omega )\beta _{j}(\omega )d\omega  \\
&&+\sum_{j=1,2}\int\limits_{-\infty }^{+\infty }i\sqrt{\frac{\kappa _{j}}{%
\pi }}[b_{j}^{\dag }(\omega )a_{j}-a_{j}^{\dag }b_{j}(\omega )]d\omega
\notag \\
&&+\int\limits_{-\infty }^{+\infty }i\sqrt{\frac{\gamma _{1}}{\pi }}[\beta
_{1}^{\dag }(\omega )\sigma _{13}-\sigma _{31}\beta _{1}(\omega )]d\omega
\notag \\
&&+\int\limits_{-\infty }^{+\infty }i\sqrt{\frac{\gamma _{2}}{\pi }}[\beta
_{2}^{\dag }(\omega )\sigma _{23}-\sigma _{32}\beta _{2}(\omega )]d\omega
\notag \\
&&+\int\limits_{-\infty }^{+\infty }i\sqrt{\frac{\gamma _{3}}{\pi }}[\beta
_{3}^{\dag }(\omega )\sigma _{24}-\sigma _{42}\beta _{2}(\omega )]d\omega ,
\notag
\end{eqnarray}%
with $\{b_{j}(\omega ),\beta _{j}(\omega )\}$ the reservoir's annihilation
operators at frequency $\omega $ and $\{\kappa _{j},\gamma _{j}\}$ the
corresponding decoherence rates. Here we have made the Markovian
approximation that the system-bath coupling coefficients are constant around
the frequencies of interest.

The master equation of the density matrix $\rho $ of the system is%
\begin{eqnarray}
\frac{d\rho }{dt} &=&i[\rho ,H_{sys}]+\kappa _{1}L[a_{1}]\rho +\kappa
_{2}L[a_{2}]\rho  \\
&&+\gamma _{1}L[\sigma _{13}]\rho +\gamma _{2}L[\sigma _{23}]\rho +\gamma
_{3}L[\sigma _{24}]\rho ,  \notag
\end{eqnarray}%
where $L[c]\rho =2c\rho c^{\dag }-c^{\dag }c\rho -\rho c^{\dag }c$ is the
Lindbladian form. The expectation value $\left\langle O\right\rangle $ of a
particular system operator $A$ thus evolves as%
\begin{equation}
\frac{d\left\langle O\right\rangle }{dt}=\mathrm{Tr}(O\frac{d\rho }{dt}%
)=\left\langle K\right\rangle ,  \label{Eq GeneralEvolution}
\end{equation}%
where
\begin{eqnarray}
K &=&i[H_{sys},O]+\kappa _{1}M[a_{1}]O+\kappa _{2}M[a_{2}]O  \label{Eq GE} \\
&&+\gamma _{1}M[\sigma _{13}]O+\gamma _{2}M[\sigma _{23}]O+\gamma
_{3}M[\sigma _{24}]O,  \notag
\end{eqnarray}%
with the M form $M[c]O=2c^{\dag }Oc-c^{\dag }cO-Oc^{\dag }c$. With the help
of Eq. (\ref{Eq GeneralEvolution}), we calculate the time evolution of all
the expectation values of the atomic operators. The evolution equations of
the population operators are%
\begin{equation}
\frac{d\left\langle \sigma _{11}\right\rangle }{dt}=\left\langle g_{1}\left(
a_{1}^{\dag }\sigma _{13}+\sigma _{31}a_{1}\right) +2\gamma _{1}\sigma
_{33}\right\rangle ,  \label{Eq M11}
\end{equation}%
\begin{eqnarray}
\frac{d\left\langle \sigma _{22}\right\rangle }{dt} &=&\left\langle
g_{2}\left( a_{2}^{\dag }\sigma _{24}+\sigma _{42}a_{2}\right) +2\gamma
_{3}\sigma _{44}\right\rangle   \label{Eq M22} \\
&&+\left\langle \Omega _{c}\left( \sigma _{23}+\sigma _{32}\right) +2\gamma
_{2}\sigma _{33}\right\rangle ,  \notag
\end{eqnarray}%
\begin{eqnarray}
\frac{d\left\langle \sigma _{33}\right\rangle }{dt} &=&-\left\langle \left\{
g_{1}\left( a_{1}^{\dag }\sigma _{13}+\sigma _{31}a_{1}\right) +2\gamma
_{1}\sigma _{33}\right\} \right\rangle  \\
&&-\left\langle \left\{ \Omega _{c}\left( \sigma _{23}+\sigma _{32}\right)
+2\gamma _{2}\sigma _{33}\right\} \right\rangle ,  \notag
\end{eqnarray}%
\begin{equation}
\frac{d\left\langle \sigma _{44}\right\rangle }{dt}=-\left\langle \left\{
g_{2}\left( a_{2}^{\dag }\sigma _{24}+\sigma _{42}a_{2}\right) +2\gamma
_{3}\sigma _{44}\right\} \right\rangle ,
\end{equation}%
while the evolution equations of the coherence operators are
\begin{equation}
\frac{d\left\langle \sigma _{12}\right\rangle }{dt}=\left\langle \Omega
_{c}\sigma _{13}\right\rangle +\left\langle g_{1}\sigma
_{32}a_{1}+g_{2}a_{2}^{\dag }\sigma _{14}\right\rangle ,
\end{equation}%
\begin{eqnarray}
\frac{d\left\langle \sigma _{13}\right\rangle }{dt} &=&\left\langle -\left(
\gamma _{1}+\gamma _{2}+i\delta \right) \sigma _{13}-\Omega _{c}\sigma
_{12}\right\rangle  \\
&&+\left\langle g_{1}\left( \sigma _{33}-\sigma _{11}\right)
a_{1}\right\rangle ,  \notag
\end{eqnarray}%
\begin{equation}
\frac{d\left\langle \sigma _{14}\right\rangle }{dt}=\left\langle -\left(
\gamma _{3}+i\Delta \right) \sigma _{14}\right\rangle +\left\langle
g_{1}\sigma _{34}a_{1}-g_{2}\sigma _{12}a_{2}\right\rangle ,
\end{equation}%
\begin{eqnarray}
\frac{d\left\langle \sigma _{32}\right\rangle }{dt} &=&\left\langle \left(
-\gamma _{1}-\gamma _{2}+i\delta \right) \sigma _{32}+\Omega _{c}\left(
\sigma _{33}-\sigma _{22}\right) \right\rangle  \\
&&+\left\langle g_{2}a_{2}^{\dag }\sigma _{34}-g_{1}a_{1}^{\dag }\sigma
_{12}\right\rangle ,  \notag
\end{eqnarray}%
\begin{equation}
\frac{d\left\langle \sigma _{24}\right\rangle }{dt}=\left\langle -\left(
\gamma _{3}+i\Delta \right) \sigma _{24}+\Omega _{c}\sigma
_{34}\right\rangle +\left\langle g_{2}\left( \sigma _{44}-\sigma
_{22}\right) a_{2}\right\rangle ,
\end{equation}%
\begin{eqnarray}
\frac{d\left\langle \sigma _{34}\right\rangle }{dt} &=&\left\langle -\left(
\gamma _{1}+\gamma _{2}+\gamma _{3}-i\delta +i\Delta \right) \sigma
_{34}-\Omega _{c}\sigma _{24}\right\rangle   \label{Eq M34} \\
&&-\left\langle \left( g_{1}a_{1}^{\dag }\sigma _{14}+g_{2}\sigma
_{32}a_{2}\right) \right\rangle .  \notag
\end{eqnarray}

When the adiabatic conditions in Eq. (\ref{Eq criteria}) are fulfilled, the
degrees of freedom of the atom follow those of the cavities. Therefore, we
set the right side of Eqs. (\ref{Eq M11})-(\ref{Eq M34}) to be zero and
represent the atomic operators by the annihilation and creation operators of
the cavities. Suppose initially the atom is prepared in its ground state $%
\left\vert 1\right\rangle $, we can choose $\left\langle \sigma
_{11}\right\rangle =1$ as a meaningful start and solve the Eqs. (\ref{Eq M11}%
)-(\ref{Eq M34}) iteratively. The stationary values up to the third order of
the iteration are%
\begin{eqnarray}
\left\langle \sigma _{11}\right\rangle _{S} &=&1-\frac{g_{1}^{2}}{\Omega
_{c}^{2}}a_{1}^{\dag }a_{1},\left\langle \sigma _{22}\right\rangle _{S}=%
\frac{g_{1}^{2}}{\Omega _{c}^{2}}a_{1}^{\dag }a_{1},  \label{Eq Sp} \\
\left\langle \sigma _{33}\right\rangle _{S} &=&\left\langle \sigma
_{44}\right\rangle _{S}=0,  \notag
\end{eqnarray}%
\begin{eqnarray}
\left\langle \sigma _{12}\right\rangle _{S} &=&-\frac{g_{1}}{\Omega _{c}}%
a_{1}+\frac{g_{1}^{3}}{\Omega _{c}^{3}}a_{1}^{\dag }a_{1}a_{1}+\frac{\left(
\gamma _{1}+\gamma _{2}+i\delta \right) g_{1}g_{2}^{2}}{\left( \gamma
_{3}+i\Delta \right) \Omega _{c}^{3}}a_{2}^{\dag }a_{1}a_{2},  \label{Eq Sc}
\\
\left\langle \sigma _{13}\right\rangle _{S} &=&-\frac{g_{1}g_{2}^{2}}{\left(
\gamma _{3}+i\Delta \right) \Omega _{c}^{2}}a_{2}^{\dag
}a_{1}a_{2},\left\langle \sigma _{14}\right\rangle _{S}=\frac{g_{1}g_{2}}{%
\left( \gamma _{3}+i\Delta \right) \Omega _{c}}a_{1}a_{2},  \notag \\
\left\langle \sigma _{32}\right\rangle _{S} &=&\left\langle \sigma
_{34}\right\rangle _{S}=0,\left\langle \sigma _{24}\right\rangle _{S}=-\frac{%
g_{1}^{2}g_{2}}{\left( \gamma _{3}+i\Delta \right) \Omega _{c}^{2}}%
a_{1}^{\dag }a_{1}a_{2}.  \notag
\end{eqnarray}%
Setting $\gamma _{3}=0$ we get Eqs. \ref{Eq S13S} and \ref{Eq S24S} from Eq. %
\ref{Eq Sc}.

In this Appendix, we have treated a relative simple case. Only few
decoherence channels are involved, while the atom-photon coupling
configuration is also very "clean". For the more complicated situations
discussed in the manuscript, this systematic method is still valid: We first
modify $H_{whole}$ according to the problems we consider and then re-derive
the evolution equations for the atomic operators; After performing the
adiabatical elimination and iteration, we can get the stationary values of
the atomic operators which contain the information of the effective
evolution of the two cavity modes.

\end{document}